\documentstyle[12pt]{article}
\input epsf.tex
\def\epsfsetgraph#1{%
   \epsfrsize=\epsfury\pspoints
   \advance\epsfrsize by-\epsflly\pspoints
   \epsftsize=\epsfurx\pspoints
   \advance\epsftsize by-\epsfllx\pspoints
   \epsfsize\epsftsize\epsfrsize
   \ifnum\epsfxsize=0 \ifnum\epsfysize=0
      \epsfxsize=\epsftsize \epsfysize=\epsfrsize
     \else\epsftmp=\epsftsize \divide\epsftmp\epsfrsize
       \epsfxsize=\epsfysize \multiply\epsfxsize\epsftmp
       \multiply\epsftmp\epsfrsize \advance\epsftsize-\epsftmp
       \epsftmp=\epsfysize
       \loop \advance\epsftsize\epsftsize \divide\epsftmp 2
       \ifnum\epsftmp>0
          \ifnum\epsftsize<\epsfrsize\else
             \advance\epsftsize-\epsfrsize \advance\epsfxsize\epsftmp \fi
       \repeat
     \fi
   \else\epsftmp=\epsfrsize \divide\epsftmp\epsftsize
     \epsfysize=\epsfxsize \multiply\epsfysize\epsftmp
     \multiply\epsftmp\epsftsize \advance\epsfrsize-\epsftmp
     \epsftmp=\epsfxsize
     \loop \advance\epsfrsize\epsfrsize \divide\epsftmp 2
     \ifnum\epsftmp>0
        \ifnum\epsfrsize<\epsftsize\else
           \advance\epsfrsize-\epsftsize \advance\epsfysize\epsftmp \fi
     \repeat
   \fi
   \ifepsfverbose\message{#1: width=\the\epsfxsize, height=\the\epsfysize}\fi
   \epsftmp=10\epsfxsize \divide\epsftmp\pspoints
   \vbox to\epsfysize{\vfil\hbox to\epsfxsize{%
      \includegraphics{#1}%
      \hfil}}%
\epsfxsize=0pt\epsfysize=0pt}%
\newcount\epsfscale    %% scaling factor
\def\epsfsize#1#2{%
  \ifnum\epsfscale=1000
  \else \epsfxsize=#1
    \divide\epsfxsize by 1000 \multiply\epsfxsize by \epsfscale
    \epsfscale=1000
  \fi}

\epsfverbosetrue

\newtoks\test
\test{800}

\newdimen\hsizexxx
\hsizexxx= \hsize
\divide\hsizexxx   by 3
\multiply\hsizexxx by 2

\def\vfigspacei{ \null\vspace{-0.8cm}}
\def\vfigspaceii{\null\vspace{-0.5cm}}

\def\vfigspacei{ \null\vspace{-0.5cm}}
\def\vfigspaceii{\null\vspace{-0.6cm}}

\def\doitii{
\message{ ******  }\message{
\the\epsfxsize     ...horizontal size after scaling   }\message{
\the\epsfysize     ...vertical size after scaling   }\message{
\the\epsftsize     ...horizontal size before scaling   }\message{
\the\epsfrsize     ...vertical size before scaling    }\message{
\the\epsfscale     ...computed scaling factor }
}

\def\figi{
\begin{figure}[t]
 \epsfxsize= \hsizexxx
 \epsfscale=700
  \centerline{\epsfbox{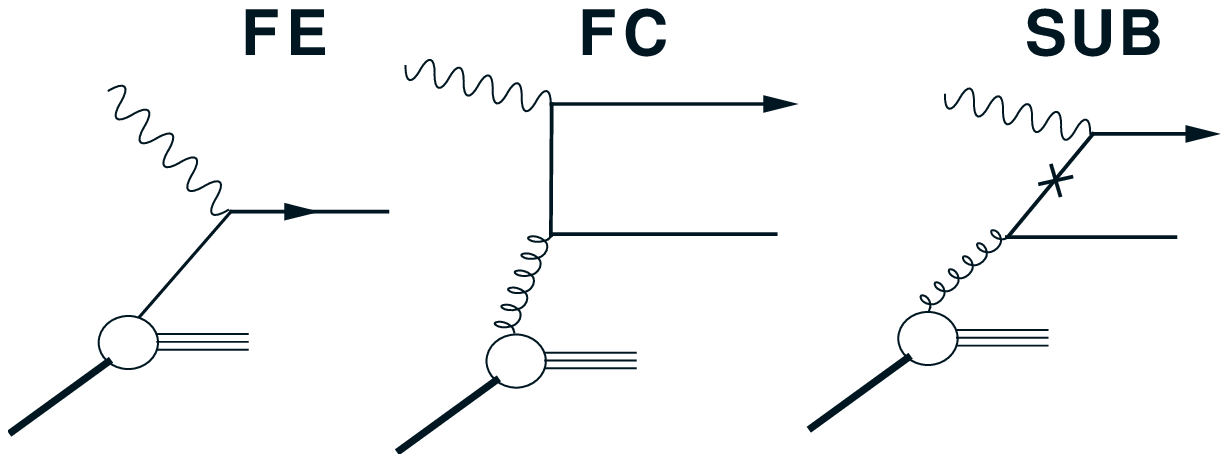}}
\vfigspacei
\doitii
      \caption{
{\protect\small
 Representative leading order diagrams for
flavor-excitation and flavor-creation production mechanisms and the overlap
between the two (which must be subtracted for consistency).
}  }
   \label{fig:}
\vfigspaceii
\end{figure}
}
\def\figii{
\begin{figure}[t]
 \epsfxsize= \hsize
 \epsfscale=460
 \epsfscale=450
   \centerline{\epsfbox{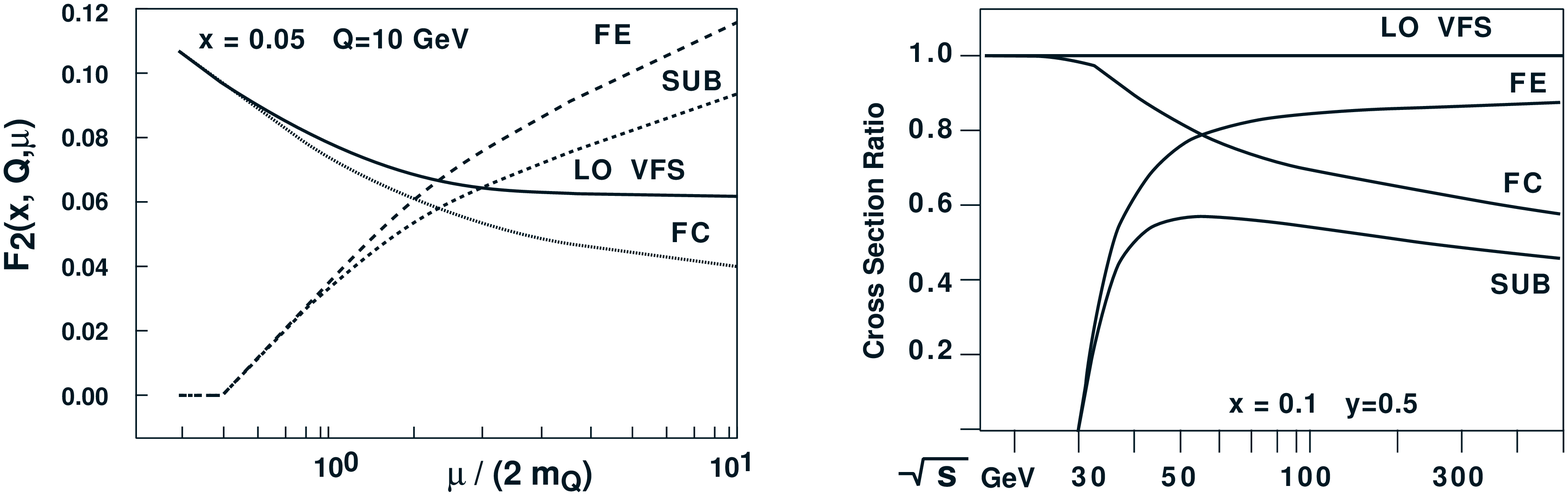}}
\vfigspacei
\doitii
      \caption{
{\protect\small
a) Scale dependence of the contributing terms to $F_2(x,Q)$ for  bottom
production.
 The factorization scale $\mu$ is shown in units of the physical scale
$2m_Q$.
 b) Relative contributions of these terms to the physical cross section
for $b$-production,
   $d\sigma/dx /dy$ at $\{ x,y\}=\{0.1,0.5\}$
   \protect{\it vs.\protect} CM energy $\protect\sqrt{s}$ in GeV
   normalized to the full LO result in the {\it variable flavor scheme} (VFS).
 } }
   \label{fig:}
\vfigspaceii
\end{figure}
}
\def\figiii{
\begin{figure}[t]
 \epsfxsize= \hsize
 \epsfscale=880
 \epsfscale=850
   \centerline{\epsfbox{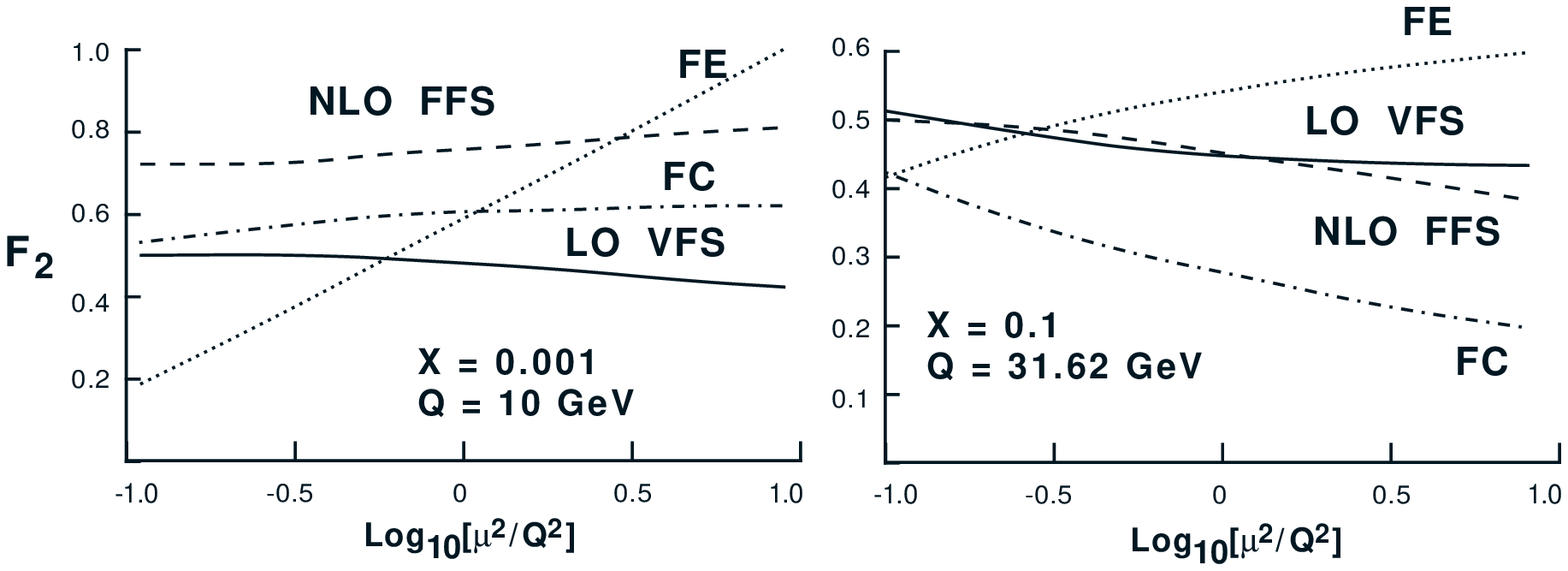}}
\vfigspacei
\doitii
      \caption{
{\protect\small
Scale dependence of the various calculations of $F_2(x,Q)$ for  charm
production on the factorization scale $\mu$.
}   }
   \label{fig:}
\vfigspaceii
\end{figure}
}
\def\figiv{
\begin{figure}[t]
 \epsfxsize= \hsizexxx
 \epsfscale=460
 \epsfscale=700
   \centerline{\epsfbox{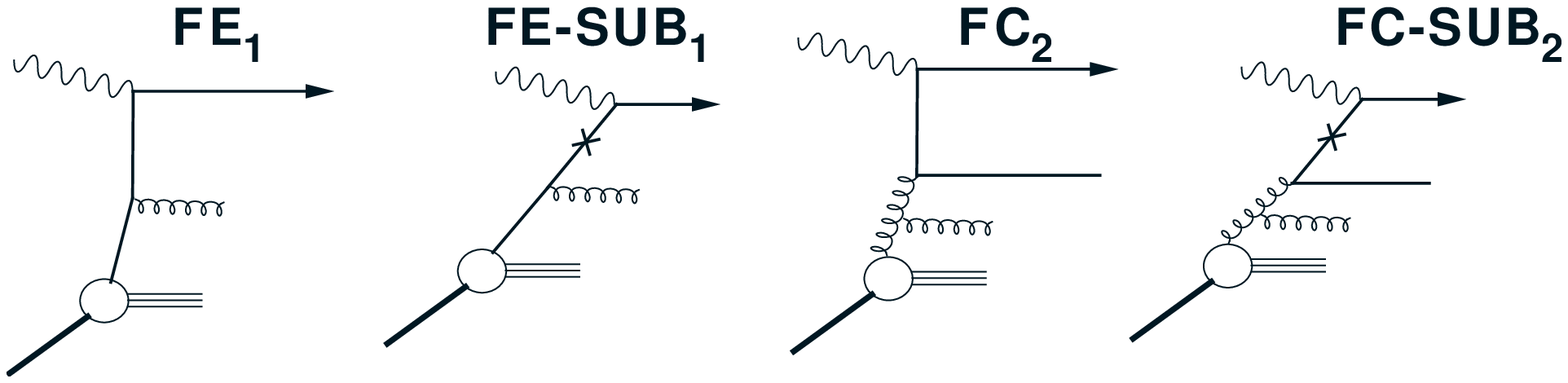}}
\vfigspacei
\doitii
      \caption{
{\protect\small
Representative next-to-leading diagrams for FE and FC production
mechanisms and their overlaps. All terms are of the same {\it numerical}
order---\protect\oas2.
}   }
   \label{fig:}
\vfigspaceii
\end{figure}
}
\def\figv{
\begin{figure}[t]
 \epsfxsize= \hsize
 \epsfscale=830
 \epsfscale=750
   \centerline{\epsfbox{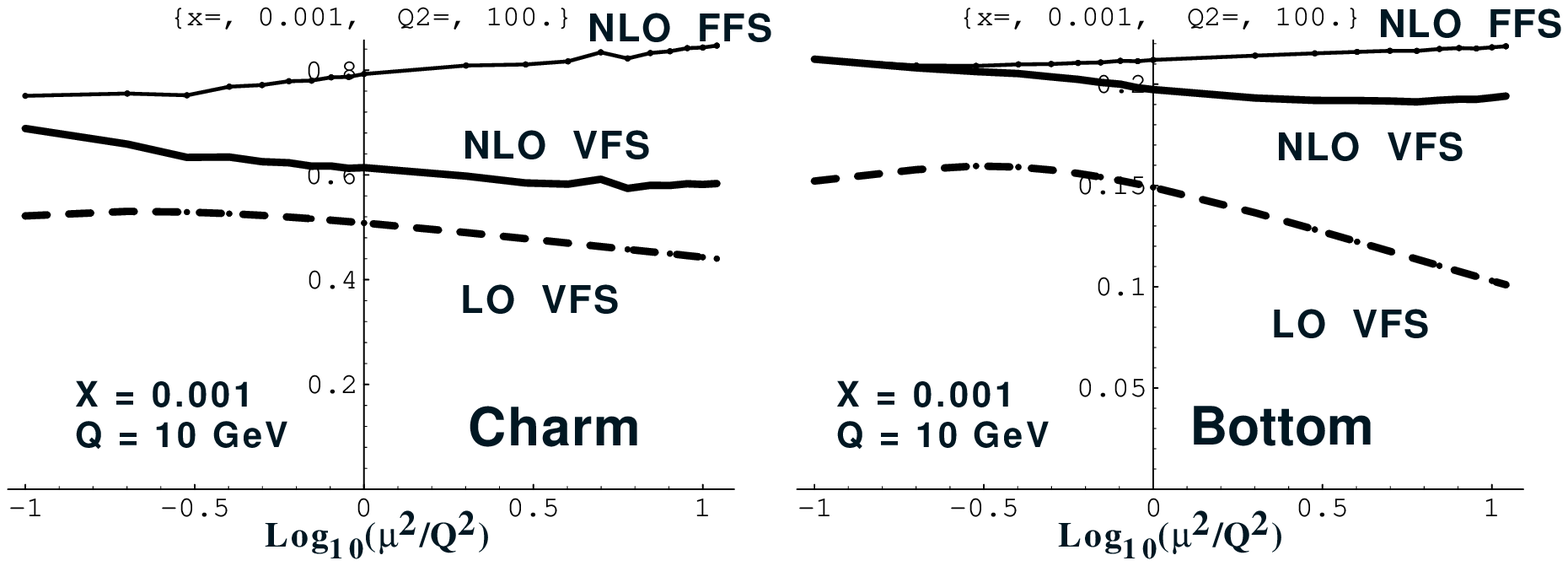}}
\vfigspacei
\doitii
      \caption{
{\protect\small
Scale dependence of the  $F_2(x,Q)$ for  charm and  bottom production
for $x=0.001$, \hbox{$Q= 10\, GeV$,} comparing the new NLO variable flavor
scheme results with existing ones.
}   }
   \label{fig:}
\vfigspaceii
\end{figure}
}
\def\fsec#1{\vspace*{-0.4cm}\section{#1}\vspace*{-0.4cm}}
\def\fsec#1{\vspace*{-0.2cm}\section{#1}\vspace*{-0.3cm}}

\epsfverbosetrue

\def\oas#1{\hbox{${\cal O}(\alpha_s^{#1})$}}
\def\gsim{\lower0.5ex\hbox{$\stackrel{>}{\sim}$}}
\def\lsim{\lower0.5ex\hbox{$\stackrel{<}{\sim}$}}
\def\alphas{\alpha_s}

\def\frac#1#2{(#1/#2)}
\def\preprint#1{}
\def\preprint#1{#1}

\begin{document}

\begin{titlepage}
\makebox[2cm]{}\\[-1in]

\preprint{
 \begin{flushright}
 \begin{tabular}{l}
 SMU-HEP/95-04\\
 hep-ph/9507295 \\
 May 1995
 \end{tabular}
 \end{flushright}
}

\vspace*{2.0cm}
\begin{center}
{\large\bf
LEPTOPRODUCTION OF HEAVY QUARKS}

\vspace{1.3cm}
Pankaj Agrawal\rlap,{${}^a$}
Fredrick I.
Olness\rlap,{${}^b$}\footnote{Presented by Fredrick Olness}
Stephan T. Riemersma\rlap,{${}^b$} Wu-Ki Tung{${}^a$}

{\em
{${}^a$}Michigan State University, East Lansing, Michigan 48824  USA \\
{${}^b$}Southern Methodist University, Dallas, Texas 75275 USA
}

\vspace{6.5cm}
\vfill

{\bf Abstract\\[5pt]}
\parbox[t]{\textwidth}
 {\small
There are presently two approaches to calculating heavy
quark production for the
deeply inelastic scattering process in  current literature.
 The conventional {\it fixed-flavor scheme} focuses on the flavor creation
mechanism and includes the heavy quark {\it only} as a final state particle
in the hard scattering cross section;   this has been  computed to
next-to-leading order--$\alphas^2$.
 The more recently formulated   {\it variable-flavor scheme}  includes,
in addition,   the flavor excitation process where the  initial state
partons of all flavors contribute above their respective threshold,
as commonly accepted for calculations of other high energy processes; this
was initially carried out to leading order--$\alphas^1$.
 We first compare and contrast these existing  calculations.
 As expected from physical grounds, the next-to-leading-order {\it
fixed-flavor scheme} calculation yields good results near threshold,
while  the leading-order
{\it variable-flavor scheme} calculation  works well for
asymptotic $Q^2$.
 The quality of the calculations in the intermediate region is dependent
upon the   $x$ and $Q^2$ values chosen.
 An accurate self-consistent QCD calculation over the entire range can be
obtained by extending the {\it variable-flavor scheme} to
next-to-leading-order.
 Recent work to carry out this calculation is described.
Preliminary numerical results of this calculation are also presented for
comparison.
 }

\vfill

\end{center}

\preprint{
 \centerline{\it to appear in the Proceedings of the XXXth Rencontres
 de Moriond}
 \centerline{\it ``QCD and High Energy Hadronic Interactions''}
 \centerline{\it Les Arcs, France, March 1995}
}

\end{titlepage}
\newpage

\fsec{Introduction \label{intro}}

The production of heavy quarks in photo-, lepto-, and hadro-production
processes have become an increasingly important subject of study both
theoretically and experimentally. The theory of heavy quark production in
perturbative Quantum Chromodynamics (QCD) is  subtle
 because of the additional scales
introduced by the quark masses ($m_c,\,m_b,\,m_t$), generically denoted
by $m_H$ in the following.\cite{dis,hq}
 Traditional  {\it
fixed-flavor scheme}
  (FFS) calculations treat  the
heavy quark mass $m_H$ as a {\em large parameter} for all ranges of the
physical momentum scale, generically denoted by\footnote{The generic
scale $Q$ can be identified with the familiar variable ``$Q$'' in deep
inelastic scattering and lepton-pair production (Drell-Yan process),
$M_{W,Z}$ in $W,Z$ production, $p_t$ in direct-photon and jet production,
\ldots\ etc.} $Q$;  hence, it includes only those hard processes initiated
by the gluon and the ``light quarks'' ($u,d,s$).\cite{lrsn,nde}  It is
expected to become unreliable at high energies when $Q\gg m_H,$ since the
presence of powers of $\ln (Q/\mu)$ and $\ln (m_H/\mu ) $ in the hard
cross section formulas will invalidate the perturbative expansion
regardless of how one chooses the renormalization and factorization scale
$\mu$.
As these ratios can indeed be quite large for charm and bottom quarks at
current and future colliders, it is not surprising that results of
existing fixed-order calculations have been under much scrutiny.
Specifically, both the large size of the next-to-leading-order (NLO)
corrections (compared to the leading-order (LO)  ones) and the substantial
dependence of the results on the (in principle, arbitrary) scale
parameter $\mu$ are suggestive symptoms of the large logarithm
problem.\cite{altarelli} In addition, quantitative comparisons with first
experimental results on charm and bottom production have not been entirely
satisfactory.\cite{cdf}

The problem can be stated in another way: in the kinematic regime where $Q
\gg m_H$, the heavy quark mass $m_H$ becomes {\em relatively} ``light.''
It is then expected to behave just like one of the familiar light quarks
-- it should be more naturally treated like another ``parton.''  This
transition from a heavy particle when $Q \lsim m_H$ to a parton when $Q \gg
m_H$ is intuitively obvious; it is implicitly accepted in general
considerations of high energy processes both within and beyond the
standard model,\cite{ehlq} and explicitly incorporated in all calculations of
parton
distribution functions (PDF's) inside the hadron\cite{mrs,cteq} -- where the
number of effective parton flavors increases each time a quark threshold is
crossed. However, as mentioned above, the conventional {\it fixed-flavor
scheme} of hard-scattering cross sections of heavy flavor production does
not incorporate this effect -- a quark with mass $m_H$ is treated as
``heavy'' for all values of $Q$, and it is never counted as a parton. Two
problems arise from this scheme: (i) the calculation clearly becomes
unnatural and unreliable when $Q \gg m_H;$ and (ii) the combined use of
PDF's calculated in a scheme with scale-dependent number
of parton flavors in conjunction with hard cross sections calculated in a
different scheme with fixed (scale-independent) number of partons is
obviously an inconsistent application of QCD.

The solution to this problem lies in formulating a consistent
renormalization and factorization scheme\cite{cwz} which explicitly implements
the above mentioned transition of a ``heavy quark'' to a ``light parton."  To
distinguish it from the FFS, we shall designate this the {\it variable
flavor scheme} (VFS).\cite{acot}  In the VFS, there is an intricate interplay
between the gluon-boson-fusion (or ``flavor creation'') and quark-scattering
(or
``flavor excitation'') production mechanisms as the typical physical
scale $Q$ varies from the threshold to the asymptotic regions.

In the following, we will review the theoretical issues of heavy quark
production relevant for present and future experiments, and present a
brief comparison of the existing calculations.  We then describe a new
three-order calculation for heavy quark production in the VFS which is
valid from the threshold region to asymptotic energies.  This new result
will allow us to improve the quantitative QCD theory of heavy flavor
lepto-production. The principles described here are also applicable to
hadro-production.
 Not mentioned here are the higher-order QCD corrections to heavy quark
production due to large logarithms associated with ``small-x" which require
resummation of an entirely different type.\cite{resum}

\fsec{Variable Flavor Scheme: From Low to High Energy \label{sec:}}

\vspace*{-0.2cm}

\figi

The intuitive notion that a ``heavy quark'' with mass becomes {\em
decoupled} from physical processes at a scale $Q$ $\ll m_H$ (thus should
not be counted as one of the ``partons''), and that the same quark becomes
an {\em active parton} at a much higher scale $Q \gg m_H$, is implemented
precisely in terms of the VFS in which the running coupling function
$\alpha_s(\mu )$ and the PDF's $f_N^H(\xi ,\mu )$
are continuous across the heavy quark threshold
$\mu_{th}=m_H$.[cwz,ColTun,lhk2]  Decoupling manifests itself in
$f_N^H(\xi ,\mu )=0$ for $\mu \leq \mu_{th}$, and the effective flavor
number $n_f$ does not include $H$ below threshold.  Above the threshold
$\mu \geq \mu_{th}=m_H$, $f_N^H(\xi ,\mu )$ satisfies the PDF
evolution equation with the usual $\overline{MS}$ evolution kernel, and
the effective flavor number $n_f$ is incremented by one to count
$H$ as one of the active  partons.\footnote{ As shown in ref.~\cite{ct}, for
these intuitively natural properties of
$\alpha_s(\mu )$ and $f_N^H(\xi ,\mu )$ to hold, the threshold $\mu_{th}$
must be chosen at $m_H$ -- not at $c \, m_H$ with some arbitrarily chosen
$c$ (such as $c=2$ or $4$).}

This scheme naturally includes both the {\it flavor excitation} and the
{\it flavor creation} production mechanism; the basic idea is illustrated
in Fig.~1 which shows one representative leading order diagram of each
kind.  The quark
initiated {\it flavor excitation} (FE) diagram contributes when the heavy
quark PDF $f_N^H(\xi ,\mu )$ is non-vanishing.  It
contains the resummed collinear logarithms to all orders, and represents
the dominant physics at
large
$Q^2/m_H^2$.  The gluon initiated {\it flavor creation} (FC) diagram
captures the correct physics at energy scales of the same order as
the quark mass $m_H$.  The {\it subtraction}
(SUB) diagram represents the overlap between the two complementary
mechanisms, and serves the dual purpose of removing the double counting
and cancelling the
collinear singularity contained in FC for large $\log(Q^2/m_H^2)$. It is
important to realize that, although the leading FE diagram is formally of
\oas0\ and the other two are of \oas1, all three are, in fact of the
{\em same numerical order} since the heavy-quark distribution function
of the FE term is effectively of \oas1\ compared to the gluon distribution
of the others. It is in this sense, results based on these
diagrams\cite{acot} are referred to as ``leading-order VFS''
calculation.\footnote{
Not included here are the quark-initiated
\oas1\ FE contributions which, for the same reason mentioned here, are
numerically of the same order as \oas2\ FC terms. They need to be
included only for the NLO VFS calculation, as will be discussed in
Sec.~4}

\figii

The main features of the VFS, highlighting the interplay between
the FE, FC production mechanisms, and the subtraction term in
the various energy ranges, are illustrated in
Fig.~2 where the separate contributions to the b-quark production
structure function $F_2$ and the production cross section at a specific
$\{x, y\}$ point
are shown as a function of the factorization scale $\mu$ and the center-
of-mass energy respectively.

First, consider Fig.~2a. At a small scale
$\mu$, the b-quark is not a constituent of the hadron, therefore the
b-quark initiated FE term vanishes. In this region,
subtraction vanishes as well so that the total $F_2$ is given completely
by the gluon initiated FC term.

As the $\mu$ scale increases, the b-quark evolves as a parton in the
hadron, and the b-quark PDF increases rapidly due to splitting of the
large number of gluons present.  This large $\mu$-dependence by itself
appears rather unnatural and may be a cause of serious concern.  However,
in the threshold region, the collinear subtraction term has precisely the
correct form to cancel the unphysical contribution and to remove this
artificial scale dependence.
 This is because in the threshold region ($\mu\simeq m_H$), the heavy quark
distribution function is approximately given by the convolution of the
gluon distribution with the
gluon-quark splitting, $f_H \propto f_g \otimes P_{g\to H}$.  This
built-in cancellation ensures that the total physical result is actually
well-approximated by the FC term (as one would expect just above
threshold), and is minimally sensitive to not only the factorization scale,
but also the choice of renormalization scheme.

In the limit of very large $\mu$ ($\mu\gg m_H$),  the large collinear logs
in FC are canceled by the subtraction term.  The difference is
``infrared safe'' in the high energy limit, and remains numerically small.
 Consequently, $F_2$ is dominated by the FE
process, whose contribution becomes dominant as the quark distribution
function evolves to an increasingly important size.

 The same physics, concerning the two complementary production mechanisms
and their relative importance as a function of relevant physics scales,
is shown in Fig.~2b for the physical cross section {\it vs.} energy over
the range from fixed-target to HERA. For this
comparison, the various terms are normalized to the full result. One can
see clearly that  $\sigma_{VFS} \simeq
\sigma_{FC}$ at low energies, and  $\sigma_{VFS} \simeq \sigma_{FE}$  in
the high energy limit.\cite{dpf92}

%\newpage

\fsec{Comparison of VFS and FFS \label{sec:}}

\figiii

A systematic comparison of the previously available VFS and FFS
calculations (LO and NLO respectively) is presented in ref.~\cite{or}.
We summarize the highlights below.
 In the NLO FFS calculation,\cite{lrsn}
 the heavy quark enters into the hard
scattering {\it only}  via the FC process.
 (For example, when considering $b$ production, the number of
{\it light} flavors would be four.)
 The calculation has been carried out to \oas2, and this represents the most
complete calculation for physical scales $Q$ not too far above $m_H$.
Problems will arise when terms containing ``collinear logarithms''
$\log^n(Q^2/m_H^2)$
$(n=1,2)$ become significant. Fortunately, $Q^2/m_H^2$ stays moderate
even for the highest energy deep
inelastic scattering experiments. Thus, as shown in Fig.~3
for a representative situation at HERA energies, the FFS result is quite
stable against variation of the (arbitrary) factorization scale--a good
indication that it should be reliable. This is in rather sharp contrast
to the case of comparable calculations of heavy quark production in
hadron colliders\cite{altarelli} where the relevant scale
$P_T^2$ can be very large compared to $m_c^2$ and $m_b^2$, leading to
well-known unstable results against choice of the scale,\cite{altarelli}
({\it e.g.}, at the Tevatron with
$P_T=50\,GeV$, $P_T^2/m_c^2\sim 10^3$).

In the VFS approach, the contribution from the FE process with initial
state c- and b-quark distributions represents the resummation of $\alpha
_s^n\,\log ^n(Q^2/m_H^2)$ terms {\it to all orders} in $n$. The
``subtraction term'' (Fig.~1) removes these collinear divergences from
the FC diagrams and eliminates double counting between the FE and FC terms.
The complete treatment of the $\alpha _s^n\,\log ^n(Q^2/m_H^2)$ terms
guarantee that the VFS results are reliable at very high energies where
$Q^2\gg m_H^2$.  In principle, the VFS approach also reproduces the FFS
results when computed to the same order of $\alpha _s$, ({\it cf.},
refs.~\cite{acot,or}).
 In practice, since the LO VFS  results
only include the relevant contributions to \oas1, these results lack the
\oas2\ finite ({\it i.e.} non-collinear
divergent) pieces of the NLO FFS.
 Hence, the $\mu$-dependence of these results is not obviously better at the
energy scale shown.
 Note, there is a considerable gap between the two calculations in the
left-hand-side of the plot; this strongly suggests the need for improvement on
existing calculations.

In contrast to the above, both the leading order quark initiated
 FE process, and the leading order gluon initiated
 FC process, by themselves, have comparably large scale
dependence  ({\it cf.}, Fig.~3) which makes them entirely
unreliable in  computing
structure functions.\cite{or}

\fsec{NLO Calculation in the Variable Flavor Scheme \label
{sec:}}

\figiv

The LO VFS and
NLO FFS calculations include different
aspects of higher order corrections to the simple FE and FC heavy quark
production mechanisms which become important in different kinematic
regions. It is clearly desirable to incorporate both these corrections in
one unified treatment which is reliable for all kinematic ranges. The
VFS, by its very nature, provides the framework to do so---if we extend
the calculation to \oas2. As mentioned before
(and explained in detail in ref.~\cite{acot}), this scheme reproduces the
FC results of the FFS to the same order near threshold, and it yields the
FE results of the naive parton model in the asymptotic region---as it
should. The extension of the VFS calculation requires calculating the
\oas1\ (heavy) quark-initiated FE and the \oas2\
FC contributions, as well as identifying the overlapping collinear
subtraction terms between the two ({\it cf}. Fig.4). The calculation of
the
\oas1\ FE results, for general vector boson couplings and
arbitrary quark masses, have recently been completed.\cite{aort}  The
\oas2\ FC contribution can be taken from the available
results of the NLO FFS calculation.\cite{lrsn} One only
needs to identify the appropriate subtraction to remove the \oas2\
collinear-singularity terms which are already resummed into the QCD-evolved
FE diagrams.

In the NLO FFS calculation of ref.~\cite{lrsn},  the collinear singularities
for the light degrees of freedom (light quarks and gluons)  have, of course,
been subtracted  using dimensional regularization. All that remains to do is to
subtract the collinear
singularities associated with the ``heavy'' quark which manifest
themselves as logarithms of $m_H$. These terms are proportional to
the second-order splitting functions $P_{gH}$ and $P_{qH}$.
 To ensure the complete removal of
double counting and precise matching between the FE
contributions and the subtraction contributions near threshold, one must
use the 2-loop evolved PDF's in calculating the FE
contributions.

Fig.~5 contains preliminary numerical results from this
calculation, plotted alongside with those already mentioned from the
earlier VFS and FFS calculations. As noted before, the $\mu $-dependence
of the LO VFS and
NLO FFS calculation compensate each other.
The NLO VFS result  (for this particular
choice of $x$ and $Q^2$) lies between these complementary calculations.
For decreasing $\mu $, the NLO VFS result
approaches the
NLO FFS, and for  increasing $\mu$ it
approaches the LO VFS. The difference
between the NLO VFS result and the
NLO FFS is larger for the lighter charm
quark than the heavier bottom quark as expected since there is more phase
space over which the charm quark PDF can evolve.

These preliminary results are encouraging, and a more complete study is in
progress.\cite{aort} The result should be a calculation that will provide
an important test of perturbative QCD when compared with the results from
HERA. Perhaps of more interest is the application of the same principles
to heavy quark production in hadron-hadron collisions. As mentioned
earlier, because the ratio $Q/m_H$ (where $Q\rightarrow p_t$) is much
larger there, the conventional FFS calculations are known to be
unreliable. The proper implementation of the VFS calculation to the same
order there is therefore potentially very useful.
 For completeness, in the small-x region the complementary small-x resummation
may also be important.
 Theoretical improvement in that front would also be needed.

\figv

%\fsec*{Acknowledgments}

The authors thank John Collins, Jack Smith, and Davison Soper for useful
discussions, and  F.O. thanks P.~Aurenche and Edmond~L.~Berger for the
conference invitation.  This work is supported by the U.S.~Department of
Energy and the National Science Foundation.

%%%%%%%%%%%%%%%%%%%%%%%%%%%%%%%%%%%%%%%%%%%%%%%%%%%%%%%%
\def\fredbibitem#1{\vspace*{-0.0cm}\bibitem{#1}\vspace*{-0.25cm}}

%\newpage

\end{document}